\documentclass[12pt,a4paper]{article}
\usepackage[utf8]{inputenc}
\usepackage{setspace}

\usepackage{amsmath, amsfonts, amssymb}
\usepackage{comment}
\usepackage{graphicx}
\usepackage{psfrag}
\usepackage[usenames,dvipsnames,svgnames,table]{xcolor}
\usepackage{enumerate}
\usepackage{tcolorbox}
\usepackage{soul}
\usepackage{hyperref}
\usepackage{subfig}
\usepackage{cite}
  \usepackage{a4wide}
\usepackage{ulem}
\usepackage{tikz}
  
  \usepackage{color}
  \definecolor{dark-gray}{gray}{0.20}
  \definecolor{gray}{gray}{0.30}
  \definecolor{light-gray}{gray}{0.80}
  \definecolor{dark-red}{rgb}{0.7,0,0}
  \definecolor{dark-green}{rgb}{0.1,0.4,0}
  \definecolor{dark-blue}{rgb}{0.3,0.3,0.7}
  \definecolor{light-blue}{rgb}{0.8,0.8,1}

\usepackage{hyperref}
\usepackage{setspace}
\hypersetup{
	colorlinks=true,
	linkcolor=dark-blue,
	citecolor=dark-red,
	urlcolor=dark-green,
	linktoc=page
}
\newcommand{\be}{\begin{equation}}
\newcommand{\ee}{\end{equation}}

\def\be{\begin{equation}}
\def\ee{\end{equation}}
\def\bea{\begin{eqnarray}}
\def\eea{\end{eqnarray}}

\renewcommand{\Im}{\text{Im}\,}

\newcommand{\dd}{\mathrm{d}}

\def\Kahler{K{\"a}hler}
\def\simleq{\; \raise0.3ex\hbox{$<$\kern-0.75em
      \raise-1.1ex\hbox{$\sim$}}\; }
   \def\simgeq{\; \raise0.3ex\hbox{$>$\kern-0.75em
      \raise-1.1ex\hbox{$\sim$}}\; }

\numberwithin{equation}{section}
\begin{document}

\begin{center}
	
	{\LARGE {\bf Conifold dynamics and axion monodromies }} \\
	\vspace{0.5cm}

	\vspace{1.5 cm} {\large M. Scalisi, P. Soler, V. Van Hemelryck and T. Van Riet }\\
	\vspace{0.5 cm} 
	
	{Institute of Theoretical Physics, KU Leuven,\\
		Celestijnenlaan 200D B-3001 Leuven, Belgium
	}

	\vspace{0.7cm} {\small \upshape\ttfamily  marco.scalisi, pablo.soler, vincent.vanhemelryck,  thomas.vanriet @ kuleuven.be
	}  \\
	
	\vspace{2cm}
	
	{\bf Abstract}
\end{center}
{
It has recently been appreciated that the conifold modulus plays an important role in string-phenomenological set-ups involving warped throats, both by imposing constraints on model building and for obtaining a 10-dimensional picture of SUSY-breaking. In this note, we point out that the stability of the conifold modulus furthermore  prevents large super-Planckian axion monodromy field ranges caused by brane-flux decay processes down warped throats. Our findings imply a significant challenge for concrete string theory embeddings of the inflationary flux-unwinding scenario.
}

\setcounter{tocdepth}{2}
\newpage

%%%%%%%%%%%%%%%%%%%%
\tableofcontents

\vspace{1cm}

%%%%%%%%%%%
\section{Introduction}
%%%%%%%%%%%
Warped throats provide some of the most interesting playgrounds for string phenomenology due to their capacity of generating exponential hierarchies between energy scales. One of their most remarkable roles arises in the context of supersymmetry (SUSY) breaking and, in particular, in the construction of de Sitter (dS) vacua~\cite{Kachru:2003aw}. Long throats can be used to suppress and keep control over SUSY breaking effects induced by anti-D3-branes to an otherwise SUSY geometry. Anti-branes naturally live at the tip of warped throats and see their tensions exponentially redshifted. Under the right circumstances, this tension may provide just enough energy to generate a net positive vacuum energy, yet be small enough to maintain perturbative control and to not destabilise the geometry. 

Despite thorough scrutiny, the viability of dS uplifts in concrete compactifications and the required delicate balance of hierarchies is still the subject of much debate (see~\cite{Danielsson:2018ztv} and~\cite{Cicoli:2018kdo} for recent discussions from complementary viewpoints). The existing controversy regarding the dS landscape in string theory has only been rising since it became part of the web of Swampland conjectures that de Sitter vacua might simply not be there at all \cite{Obied:2018sgi, Ooguri:2018wrx, Danielsson:2018ztv} (see \cite{Hebecker:2018vxz, Andriot:2018wzk, Danielsson:2018qpa, Garg:2018reu, Garg:2018zdg} for related ideas) or have dramatic shorter lifespans than assumed sofar \cite{Bedroya:2019snp}.\footnote{The idea of the Swampland was first contemplated in \cite{Vafa:2005ui}  and reviewed recently in \cite{Palti:2019pca}.}

Part of the recent efforts have been devoted to analysing the effects that light geometric modes localised down warped throats can have on the effective theory. For instance, local KK modes cannot really be decoupled from the 4d effective field theory (EFT) although their couplings might be rather harmless and they could just behave as spectators \cite{Blumenhagen:2019qcg}. Of particular importance is the so-called {\it conifold modulus}, which is the local complex structure modulus of the throat first studied by Douglas and collaborators in \cite{Douglas:2007tu}. This field has a mass scale that is not as light as the K\"ahler moduli but is much lighter than the complex structure moduli of the bulk Calabi-Yau (CY). It has been recently appreciated that it might affect the stability of the KKLT scenario if flux numbers are too low \cite{Bena:2018fqc, Dudas:2019pls, Bena:2019sxm, Blumenhagen:2019qcg}. At the same time, it is claimed to be crucial for reaching a detailed 10d understanding of how anti-brane uplifting could work \cite{Randall:2019ent}.

Moduli stabilisation scenarios with long warped throats are not only useful in the study of dS vacua. As pointed out in \cite{Gautason:2016cyp, DiazDorronsoro:2017qre}, such set-ups can incorporate models of {\it axion monodromy} with large (in principle transplanckian) field ranges (see also~\cite{McAllister:2008hb,Retolaza:2015sta,Hebecker:2018yxs,Kim:2018vgz}  These have potential implications for phenomenology, in particular for the construction of models of large field inflation. More importantly, they represents an explicit framework where conceptual issues associated to transplanckian field displacements can be analysed in detail.

The Swampland distance conjecture (SDC) \cite{Ooguri:2006in} provides in fact a concrete argument why EFTs compatible with quantum gravity should not allow for parametrically large field displacements $\Delta \phi \gg M_\text{Pl}$ within their regime of validity. The typical obstruction would come from an infinite tower of states becoming exponentially light along a trajectory in field space, which leads to a decrease of the quantum gravity cut-off  \cite{Dvali:2007wp,Dvali:2007hz}. Detailed investigations in string theory have in fact verified this statement repeatdly \cite{Grimm:2018ohb,Blumenhagen:2018nts,Lee:2018urn,Lee:2018spm,Grimm:2018cpv,Gonzalo:2018guu,Corvilain:2018lgw,Font:2019cxq,Lee:2019wij,Gendler:2020dfp}, while others have failed to find fully trustworthy models where field ranges can extend parametrically beyond the Planck range. Models of axion monodromy are arguably the set-ups where the Swampland distance conjecture constraints are most challenging to understand and are still the subject of debate~\cite{Baume:2016psm,Valenzuela:2016yny,Blumenhagen:2017cxt,Blumenhagen:2018hsh,Buratti:2018xjt,Scalisi:2018eaz}.

In this note, we focus on the axion monodromy scenario of~\cite{Gautason:2016cyp, DiazDorronsoro:2017qre} in order to concretely examine the obstructions arising in warped throats when one tries to engineer super-Planckian excursions. We will in fact argue that the conifold modulus plays a crucial role in preventing parametrically large field displacement, when supersymmetry is broken, within the EFT. Our analysis is similar in several aspects to the one of~\cite{Bena:2018fqc, Dudas:2019pls, Bena:2019sxm, Blumenhagen:2019qcg} but incorporates the interplay of the conifold modulus with the axion-like field parametrising the process of brane-flux annihilation of~\cite{Kachru:2002gs}. We find that a large number of monodromies, and hence a large field displacement, cannot be achieved without destabilising the geometry along the direction parametrised by the conifold modulus. Our analysis is independent of the mechanism of volume stabilisation, and is hence applicable to diverse scenarios such as KKLT~\cite{Kachru:2003aw} or the large volume scenario~\cite{Balasubramanian:2005zx}. While some particular set-ups may provide other means by which transplanckian distances are censored, the conifold destabilisation mechanism we study is universal.

The rest of this note is organised as follows. In section \ref{Sec:moduli} we introduce two fields that play a crucial role in our analysis; the conifold modulus $S$ and an open-string modulus $\psi$ that mediates brane-flux transitions, which will be our axion-like scalar that undergoes monodromies. Since the brane-flux decay process inside compact CYs is crucial for our analysis we treat this separately in section \ref{Sec:5branes}. Here we discuss both the decay of NSNS and RR flux, and whether the stability of the conifold modulus leads to stronger constraints on meta-stable dS uplifts than previously considered. Our computations indicate that this is not the case. In section \ref{Sec:monodromies} we finally discuss the allowed field ranges for the axion monodromy and demonstrate that parametric super-Planckian displacements in the axion direction are strongly constrained.

\section{The conifold and brane-flux moduli} \label{Sec:moduli}
In this section, we introduce the basic ingredients required for our later analysis of instabilities of warped throats associated to conifold singularities. Our main fields of interest are the complex structure (cs) modulus of the conifold ({\textit{aka} conifold or $S$-modulus) and the open string scalar field $\psi$ that parametrises KPV brane-flux transitions~\cite{Kachru:2002gs}. As many of the ingredients are well known, we will be rather concise and refer to the appropriate literature for further details.

\subsection{The conifold modulus}
\label{sec:conifold_modulus}
We begin by introducing the conifold modulus and its dynamics. These were first discussed in~\cite{Douglas:2007tu,Douglas:2008jx}, and recently revisited in~\cite{Bena:2018fqc, Dudas:2019pls, Bena:2019sxm, Blumenhagen:2019qcg} (see also~\cite{Randall:2019ent} for an alternative perspective). We follow closely the discussion and notation of~\cite{Blumenhagen:2019qcg}. 

The Klebanov-Strassler (KS) solution~\cite{Klebanov:2000hb} features a 6d non-compact conformal Calabi-Yau metric and a 4d Minkowski part (in 10d string frame):
\begin{equation}
\label{eq:normal10D_warped_metric}
	\dd s_{10}^2 = e^{2A}\dd s_4^2 + e^{-2A} \dd s_6^2\,.
\end{equation}
The `internal' 6d space, whose metric is explicitly known, corresponds to a cone over a 5D space that is topologically $S^3\times S^2$. The warp factor $e^{2A}$ is a function of the radial coordinate $\tau$. At the tip of the throat corresponding to $\tau\to0$ (deep in the IR in the holographic language), the $S^2$ shrinks to zero size while the $S^3$ remains finite, supported by $M$ units of RR flux. The finite size of the $S^3$ in the unwarped metric $\dd s^2_6$ is controlled by a modulus $S$ such that 
$\dd s^2_6 = |S|^{3/2}\dd\tilde{s}^2_6$, where $\dd \tilde{s}^2_6$ does not depend on $S$. 
 \begin{comment}
\begin{equation}
	\dd s_6^2 = \frac{|S|^{\frac{2}{3}}}{2} \mathcal{K}(\tau) \left(\frac{\left[\dd \tau^2 + (g^5)^2\right]}{3\mathcal{K}^3(\tau)}+ \sinh^2(\frac{\tau}{2})\left[(g^3)^2+(g^4)^2\right]+\cosh^2(\frac{\tau}{2})\left[(g^1)^2+(g^2)^2\right]\right).
\end{equation}
Here $\mathcal{K}$ is a function whose details we do not need.
\end{comment}
However the conformal factor $e^{-2A}$ of the full metric is not independent of $S$. Its on-shell value can be written as
\begin{equation}\label{eq:warp_factor}
	e^{-4A(\tau)} = 2^{2/3} \frac{(\alpha' g_s M)^2}{|S|^{\frac{4}{3}}}\mathcal{I}(\tau)\,,
\end{equation} 
where $\mathcal{I}(\tau)$ is a known function of $\tau$ whose details we will not discuss. Notice from these expressions that the $S$-dependence in the warp factor exactly cancels the $S$-dependence in the 6d metric. Hence, once warping is taken into account the physical size of the $S^3$ at the tip of the throat is independent of $S$.\footnote{Caveats relevant to the validity of these remarks off-shell will be discussed momentarily.}

As a next step one imagines, like GKP \cite{Giddings:2001yu}, that the KS solution is a valid local description of the geometry near the deformed conifold singularity of a compact CY space. 
In our current understanding the $S$-field does not measure the size of the deformed $S^3$ (the so-called A-cycle), but rather the size of its dual B-cycle that extends towards the bulk CY. In the limit $S\rightarrow 0$ the B-cycle, and the throat, become infinitely long. 

It is convenient to rescale the conifold modulus into a dimensionless field $Z$ as 
\begin{equation}
\label{eq:S_to_Z}
	S = \alpha'^{\frac{3}{2}} \; g_s^{\frac{3}{4}} \; \mathcal{V}_w^{\frac{1}{2}} \; Z\,,
\end{equation}
where the warped volume in 10d Einstein frame is given by
\be\label{eq:warped_volume}
\mathcal{V}_w=\frac{1}{g_s^{3/2}(\alpha')^3}\int \dd^6y\,e^{-4A}\,\sqrt{g_6}\,,
\ee
defined using the unwarped internal metric $\dd s_6^2=(g_6)_{mn} \dd y^m\, \dd y^n$.
The warped volume determines in turn the four-dimensional Planck mass as
\begin{equation}
\label{eq:Planck_Mass_definition}
	M^2_\text{Pl} = \frac{M_s^2 \mathcal{V}_w}{\sqrt{g_s}}\,, \qquad \text{with} \qquad M_s^2 = \frac{1}{2\pi \alpha'}.
\end{equation}

The dynamics of the conifold modulus were analysed in~\cite{Douglas:2007tu,Douglas:2008jx}, which argued that it is controlled by an F-term scalar potential $V\equiv M_\text{Pl}^4 e^\mathcal{K} (|D\mathcal{W}|^2-3|\mathcal{W}|^2)$ described, for small values of the conifold modulus ($Z \ll 1$ near its minimum), by the \Kahler- and super-potential 
\begin{align}
\label{eq:kahlerpot}
	\mathcal{K} &= -3 \log\left(-i(\rho-\bar{\rho}) -9 c' g_s M^2 \frac{|Z|^{2/3}}{12}\right)-\log\left(\frac{2}{g_s}\right)\,,\\
\mathcal{W} &= -\frac{M}{2\pi i}Z \left(\log Z -1 \right) + \frac{i}{g_s}K Z\,.
\label{eq:fluxpot}
\end{align}
Here $\Im \rho = \sigma = \mathcal{V}_w^{\frac{2}{3}}$, $K$ is the $H_3$ flux quantum along the B-cycle, and $c'\approx 1.18$ a numerical factor~\cite{Bena:2018fqc,Douglas:2007tu} (we follow closely the conventions of \cite{Blumenhagen:2019qcg}). To lowest order in the large volume and small $Z$ expansion the scalar potential is given by
\begin{equation}
\label{eq:KS_potential}
	V_\text{KS}= M_\text{Pl}^4 \frac{|Z|^\frac{4}{3}}{2c' M^2 \sigma^2}\left|\frac{M}{2\pi}\log Z + \frac{K}{g_s}\right|^2.
\end{equation}

Some comments are in order. The specific $Z$ (or $S$)-dependence of this scalar potential should be taken with caution.\footnote{The same remarks apply to the anti-D3-brane potential presented next. We are grateful to Arthur Hebecker for pointing us to these issues.} Strictly speaking, the equations presented above  are to be trusted near the KS solution only. In fact, the $S$-dependence of the warp factor given in~\eqref{eq:warp_factor} is an on-shell relation. The off-shell version of this equation is not well understood, and therefore the potential for $Z$ is only known close to the KS vacuum. Interestingly, reference \cite{Randall:2019ent}  provides some complementary evidence that the form of the scalar potentials used here does make sense. This comes from an analysis of SUSY breaking mediation from the IR to the UV of a 5d warped throat. In this picture the 5D radion plays a crucial role~\cite{Brummer:2005sh}. A study of the Goldberger-Wise radion potential gives a result very close to the actual form derived in~\cite{Douglas:2007tu}. We therefore take some confidence in the potential~\eqref{eq:KS_potential} and proceed as in~\cite{Bena:2018fqc, Dudas:2019pls, Bena:2019sxm, Blumenhagen:2019qcg}, keeping in mind that a more detailed understanding of the conifold modulus could lead to modifications. These could affect quantitatively the results of later sections, but we expect that our methods and qualitative conclusions will hold up.

\subsection{SUSY breaking and the conifold instability} \label{sec:dSstability}

References~\cite{Bena:2018fqc,Blumenhagen:2019qcg,Dudas:2019pls} investigated whether the conifold modulus could trigger an instability when $p$ anti-D3-branes are added and SUSY is broken. The intuition behind this is that warping effects render this field significantly lighter than the bulk complex structure moduli, which are stabilised at high energy scales.

Anti-D3-branes naturally sit at the tip of the throat $\tau \rightarrow 0$, i.e. in the deep IR. In this regime they see their tension $\mu_3=(2\pi)^{-3} (\alpha')^{-2}g_s^{-1}$ effectively warped down and contribute to the potential energy as~\cite{Kachru:2003aw,Kachru:2003sx}

\begin{equation}
	\label{eq:full_antiD3_pot}
	V_{\overline{\text{D3}}} = 2p\, \frac{\mu_3 \,e^{4A(0)}}{g_s}=\frac{M_\text{Pl}^4}{2\pi} \frac{2 p \; e^{4A(0)}}{\sigma^3} = \frac{M_\text{Pl}^4}{2\pi} \frac{c'' p |Z|^{\frac{4}{3}}}{g_s M^2\sigma^2}\,
\end{equation}
where $c'' = 2^{\frac{1}{3}}/\mathcal{I}(0) \approx 1.75$ and we have used eq.~\eqref{eq:Planck_Mass_definition} and eqs.~\eqref{eq:warp_factor}-\eqref{eq:S_to_Z} in the second and third equality respectively.

The total potential, given by the combination of the flux induced component~\eqref{eq:KS_potential} and the anti-brane contribution~\eqref{eq:full_antiD3_pot}, still allows for a stable minumum of the conifold modulus. The on-shell value of $Z$ is given by~\cite{Bena:2018fqc}
\begin{equation}
\label{eq:minumum_D3}
	|Z_0|=\exp\left(-\frac{2\pi}{g_s}\frac{K}{M}-\frac{3}{4}+\sqrt{\frac{9}{16}-\frac{4\pi p}{g_sM^2}c'c''}\right)\,.
\end{equation}
This minimum for $|Z|$ obviously disappears when the argument of the square-root becomes negative, which indicates a runaway behaviour. The condition for stability is
\be\label{eq:S-constraint}
\frac{\sqrt{g_s} M}{\sqrt p} \gtrsim6.8\,.
\ee
Hence at weak coupling, $M$ needs to be large enough and one can worry about the constraints set by the RR tadpole, which is bounded by the maximal D3 charge available in the manifold \cite{Bena:2018fqc}.

This result is very interesting, but whether the actual inequality is giving new constraints can be debated. First of all, the known constraint against perturbative brane-flux decay already assumes~\cite{Kachru:2002gs}\footnote{The KPV bound on $p/M$ is strictly speaking only valid in a probe regime for NS5-branes which is not attainable in weakly coupled string theory. However non-trivial support for it came from \cite{Armas:2018rsy} who verified the bound in the supergravity regime. For that computation $p$ had to be larger than one, also because otherwise brane polarisation, the process underlying brane-flux decay, is not properly understood. Nonetheless, for applications in string phenomenology, the limit $p\to 1$ is usually taken. While the working assumption is that the bound on the ratio $p/M$ derived for $p>1$ will be valid when $p=1$, one should bear in mind that this approximation is not yet properly understood.}
\be\label{KPV}
\frac{p}{M}<0.08\,,
\ee
whereas keeping the size of the $S^3$ at the tip of the throat well above the string length scale requires
\be\label{eq:size}
g_sM\gg 1\,.
\ee
Together, conditions~\eqref{KPV} and~\eqref{eq:size} demand $\sqrt{g_s}M/\sqrt{p}\gg 1$, in line with the requirement~\eqref{eq:S-constraint} arising from the stability of $Z$. 
Additionally, in order to treat the anti-D3-brane as a probe on the $S^3$ one needs
\be
\frac{p}{g_sM^2}\ll1\,,
\ee
which takes again the same form as~\eqref{eq:S-constraint}. 

Nonetheless, the fact that the conifold modulus is much lighter than the bulk cs moduli but still heavier than the K\"ahler modulus is something to keep in mind when writing down effective field theories in set-ups with warped throats. 

\subsection{The brane-flux transition modulus}\label{sec:brane-flux}

The lightness of the conifold modulus can be roughly understood from it living down the throat. This makes one naturally wonder whether it can be of the same mass scale as other fields that are localised in the throat. One such field mediates brane-flux transitions and was first described in \cite{Kachru:2002gs}. This scalar effectively corresponds to the position $\psi$ of a spherical NS5-brane with $p$ units of worldvolume flux that wraps a contractible 2-cycle on the $S^3$. One can easily show that such an NS5-brane behaves identically to $p$ anti-D3-branes which polarise in the flux-background at the south pole of $S^3$. Moving to the north pole causes brane-flux annihilation and a decrease of the anti-brane charge (and thus of the positive energy). There the description of the NS5 can be given in terms of  $M-p$ D3-branes. This set-up is depicted in Figure~\ref{decay}.

\begin{figure}[ht!]
	\begin{center}
		\includegraphics[scale=0.33]{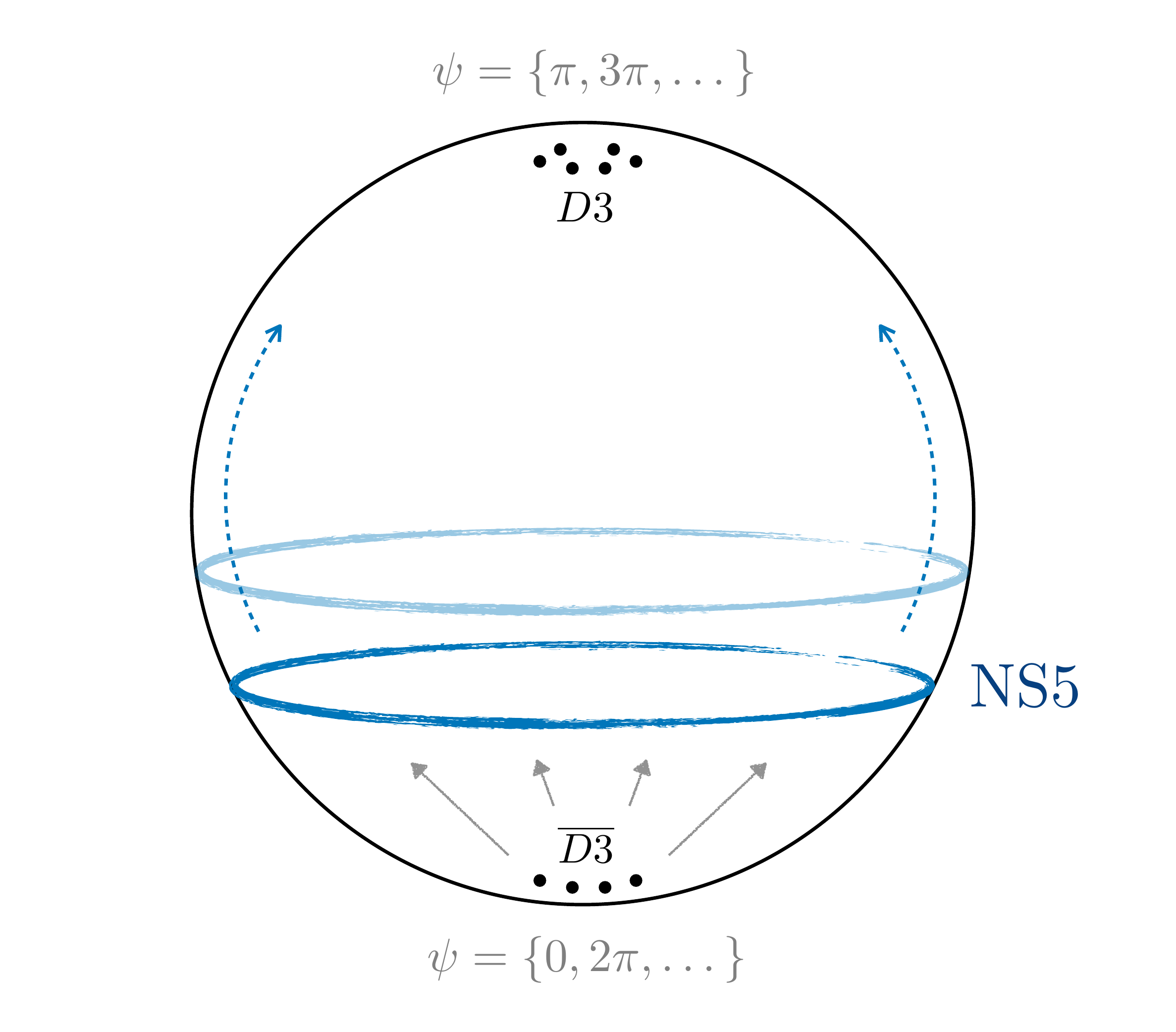}
		\vspace*{0.2cm}\caption{A schematic illustration of the brane-flux annihilation process. A number $p$ of anti-D3-branes, localised at the south pole, polarise into an NS5-brane wrapping a contractible cycle along the $S^3$. The axionic field $\psi$ describes the position of the NS5-brane, with a periodicity of $2\pi$. The south pole is at $\psi=\{0,2\pi,\dots\}$, while the north pole is at $\psi=\{\pi,3\pi,\dots\}$. The process ends when all the anti-brane charge is neutralised by the fluxes at the north pole, where the NS5 can be described in terms of $M-p$ D3-branes.}
		\label{decay}
	\end{center}
\end{figure}

It was shown by a probe computation \cite{Kachru:2002gs} that, as long as the inequality eq.~\eqref{KPV} holds, the system settles down in a meta-stable vacuum, where the NS5 behaves very much like $p$ anti-D3-branes thus still fulfilling its role as SUSY-breaking source and potential uplifter of AdS vacua~\cite{Kachru:2003aw}.  However, this meta-stable vacuum can decay through tunnelling to a configuration where the NS5 pinches and describes $M-p$ D3-branes, which do not break SUSY. This process therefore describes an ``open string'' decay channel to a SUSY vacuum by means of brane-flux annihilation. In fact, this decay along the $\psi$-direction was claimed to be the dominant decay channel in KKLT, leading to more rapid decays than the volume decompactificaton~\cite{Frey:2003dm} which is the usual focus. The picture to keep in mind is that one out of the $K$ units of NSNS 3-form flux is removed in this process and materialises into $M$ actual D3-branes.\footnote{Recall that the induced D3 charge is $KM$.} These $M$ D3-branes annihilate with the $p$ anti-D3-branes leaving $M-p$ D3-branes in the SUSY vacuum.

One can however go beyond the regime imposed by eq.~\eqref{KPV} and instead consider
\be
\frac{p}{M}\gg 1\,.
\ee
This regime is the one typical of the unwinding mechanism studied by \cite{Gautason:2016cyp, DiazDorronsoro:2017qre} and can be achieved in a probe regime. It corresponds to the periodic motion of the NS5-brane from the south-pole to the north-pole of the $S^3$ and vice versa. In this context, the position $\psi$ of the NS5 can be identified as an axionic variable. After a period of $2\pi$, it describes in fact the same position of the NS5-brane but with a lower anti-brane charge. Therefore, as long as we have positive energy, this process can potentially describe a period of cosmic acceleration (the possibility to realise inflation was contemplated in \cite{Gautason:2016cyp, DiazDorronsoro:2017qre}).  

In section \ref{Sec:5branes} we investigate the interplay between the conifold modulus and the $\psi$-modulus in the KPV regime (small $p/M$) for the sake of meta-stable dS uplifting. In section \ref{Sec:monodromies} we study the opposite regime (large $p/M$) for the sake of axion-monodromy field displacements. For the remainder of this paper, we can forget about the concrete microscopic meaning of the brane-flux annihilation process and just focus on the $\psi$-field from a 4d viewpoint. Our aim is to investigate whether this field $\psi$, since it also lives at the bottom of the throat, interacts non-trivially with the $S$-field and whether new non-trivial constraints arise. 

\section{Uplifting 5-brane runaway}\label{Sec:5branes}

In what follows we combine the effects of brane-flux decay and conifold modulus destabilisation. That is, we repeat the stability analysis of section~\ref{sec:dSstability} while allowing the $p$ anti-D3-branes to puff into an NS5-brane carrying $p$ units of anti-D3 charge as explained in~\ref{sec:brane-flux}. Following the steps laid out in~\cite{Kachru:2002gs}, one can see that the potential~\eqref{eq:full_antiD3_pot} generalises to
\begin{equation}
\label{eq:full_NS5_pot}
V_\text{NS5} = \frac{M_\text{Pl}^4}{2\pi} \frac{M e^{4A}}{\pi\sigma^3}v_\text{NS5}(\psi) = \frac{M_\text{Pl}^4}{2\pi} \frac{c''  |Z|^{\frac{4}{3}}}{2\pi g_s M\sigma^2} v_\text{NS5}(\psi),
\end{equation}
where we define 
\begin{equation}
\label{eq:vns5andU}
v_\text{NS5}(\psi) = \sqrt{b_0^4 \sin^4 \psi + U^2(\psi)} + U(\psi), \quad U(\psi) = \frac{\pi p}{M} - \psi + \frac{1}{2} \sin 2\psi
\end{equation}
with $b_0^2\approx 0.9$. Note that for $\psi=0$ the NS5-brane potential \eqref{eq:full_NS5_pot} reduces to the anti-D3-brane potential \eqref{eq:full_antiD3_pot} as it should. 

Again, the dependence on the dimensionless conifold $Z$-modulus arises through warping, c.f. eq.~\eqref{eq:warp_factor}. This dependence is hence exactly the same as for the anti-D3-brane. It is therefore not too difficult to check that the conifold modulus is stabilised, similarly to~\eqref{eq:minumum_D3}, at
\begin{equation}
\label{eq:minimumZ}
	|Z|_0 =  \exp\left(-\frac{2\pi K}{g_s M} - \frac{3}{4} + \sqrt{\frac{9}{16} - \frac{2c'c''}{g_s M}v_\text{NS5}(\psi_0)}\right)\,.
\end{equation}

The position modulus $\psi$ is stabilised exactly as in the original KPV analysis~\cite{Kachru:2002gs} at some value $\psi_0$. It is well known that a (meta-)stable minimum can be obtained only if the constraint~$p/M < 0.08$ is satisfied \cite{Kachru:2002gs}. When the dynamics of $Z$ are taken into account, one has to furthermore impose that~\eqref{eq:minimumZ} yields a meaningful result, i.e. that
\begin{equation}
\label{eq:vns5_bound}
\frac{32 c' c''}{9}\frac{v_\text{NS5}(\psi_0)}{g_s M} \approx 7.3 \,\frac{v_\text{NS5}(\psi_0)}{g_s M} <1.
\end{equation}
This is the generalisation of eq.~\eqref{eq:S-constraint}, which can be recovered by replacing $v_\text{NS5}(\psi_0)\to v_\text{NS5}(0)$. The dynamics of the NS5-brane relax the constraint~\eqref{eq:S-constraint} on the minimal allowed value of $g_s M$ (as a function of $p/M$). In the situation where the $\psi$-modulus is stabilised as in KPV when $p/M < 0.08$ \cite{Kachru:2002gs}, the bound found by \cite{Bena:2018fqc} is changed as shown in Figure~\ref{fig:gsM_min_plot}.

\begin{figure}[h]
\centering
	\begin{tikzpicture}
    	\node[anchor=south west,inner sep=0] (image) at (0,0) {\includegraphics[width=0.6\textwidth]{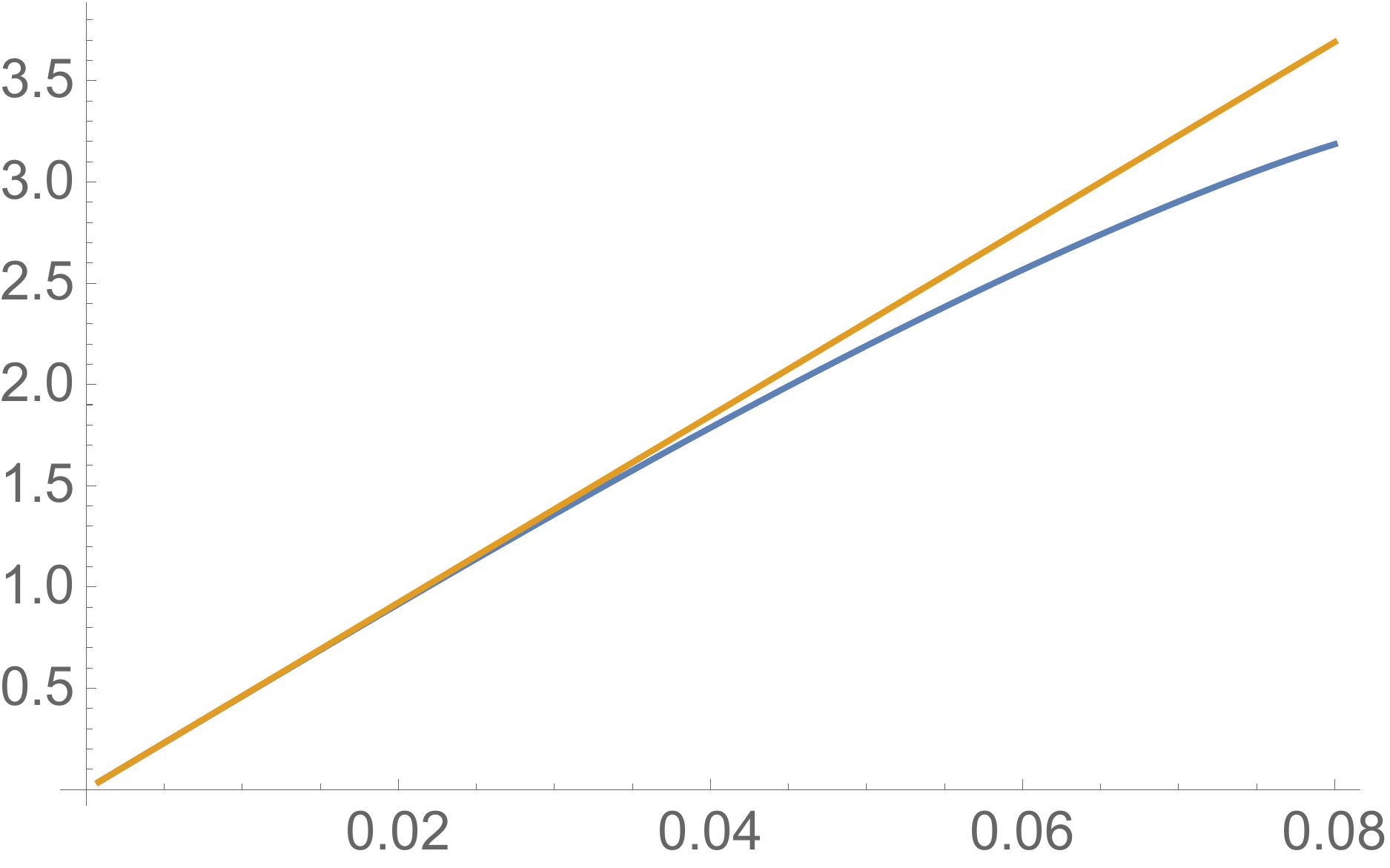}};
    \begin{scope}[
        x={(image.south east)},
        y={(image.north west)}
    ]
        \node [black] at (-0.1,1) {\large $g_s M|_\text{min}$};
        \node [black] at (1.05,0.05) {\large $\frac{p}{M}$};
    \end{scope}Runaway
\end{tikzpicture}
\caption{The minimal value of $g_s M$ allowed for avoiding the conifold instability is plotted as a function of $p/M$. The yellow line represents the bound for anti-D3-branes~\eqref{eq:S-constraint} as considered in \cite{Bena:2018fqc}, whereas the blue line represents the bound for an NS5-brane. The NS5-brane relaxes the bound, but only slightly.}
\label{fig:gsM_min_plot}
\end{figure}

In the Appendix we furthermore compare the masses of the $Z$- and $\psi$-moduli and find that for typical parameters of the KKLT scenario, the masses squared are within two orders of magnitude from each other, $\psi$ being the heavier field. This estimation is consistent with the size of the correction due to the NS5 polarisation being small. 

\subsubsection*{RR flux decay: D5 polarisation}

In a similar spirit one could worry about the dual D5 polarisation channel in which the anti-D3 charge is transferred via a polarised D5 brane moving along the B-cycle. In a non-compact model this channel is essentially irrelevant since the B-cycle extends towards the non-compact direction~\cite{Bena:2012vz}. Things might be more involved in a compact model. Indeed, in that case, the $Z$-modulus relates to the size of the B-cycle over which the D5 moves, so its dynamics will affect the brane-flux decay channel via D5 branes. Nevertheless, the expectation is that, as long as the bulk is significantly larger than the throat, this channel will be highly suppressed.\footnote{Notice that the KPV stability constraint can be written as $0.08\gtrsim p/M\sim g_sp/L_A^2$, where we used that the linear size of the A-cycle (in string units) is $L_A\sim\sqrt{g_sM}$. The factor of $L_A^{-2}$ in this bound corresponds to the 2-volume wrapped by the NS5. Similarly, one expects that decay via D5-branes  would be suppressed by some 2-volume representative of the B-cycle size $L_B^{-2}$. This size is controlled by the $H_3$ flux $L_B\sim\sqrt{K}$~\cite{DeWolfe:2004qx}, and so D5 polarisation is expected to be subdominant as long as long as there is a hierarchy $K\gg \sqrt{g_sM}$. Note however that such a hierarchy may be difficult to achieve due to tadpole constrains, e.g. in KKLT set-ups~\cite{Carta:2019rhx}.} This is in line with the general argument given in~\cite{Lanza:2019xxg} that one cannot have both relevant NSNS and RR decay channels within the same 4d EFT. The reason is that the tension of the bubble walls for one of the two processes will always be above the cut-off of the theory. This argument might in principle fail in the presence of warping, where scales get mixed up and there is never a strict separation of scales since local KK modes in the throat do not decouple from the EFT~\cite{Blumenhagen:2019qcg}.  However, a D5-brane mediating flux decay along the B-cycle would have to move into the bulk, where the effects of warping would fade. A study of the stability properties of this channel taking into account the conifold dynamics and limits set by tadpole constraints would be interesting, but it ultimately requires an explicit construction of the finite B-cycle and its embedding into a compact bulk and is outside of the scope of this note.

\section{Axion monodromies and field range bounds}\label{Sec:monodromies}
As described, the dynamics of the conifold modulus impose constraints on the parameters of (meta-)stable compactifications with warped throats and anti-branes. But these are not much more restrictive than similar bounds previously considered and can potentially be satisfied in explicit models if the landscape of string compactifications is rich enough. The main focus of this note is to explore constraints on a more ambitious set-up, where brane-flux annihilation of a large number of anti-D3-branes drives a potential of the axion monodromy type, with potential applications to inflation.

\subsection{General considerations}
Perturbatively stable vacua require a modest amount of SUSY breaking anti-branes as follows from the KPV bound (\ref{KPV}) that insists on small $p/M\lesssim 0.08$ to prevent brane-flux decay. It was speculated shortly after the KPV analysis that a larger amount of SUSY breaking, with $p/M\sim \mathcal{O}(1)$, might be useful for inflationary purposes~\cite{DeWolfe:2004qx}. Unfortunately it is difficult to make this suggestion precise because it is on the edge of trustability of the approximations made. Interestingly, as pointed out in~\cite{Gautason:2016cyp} and~\cite{DiazDorronsoro:2017qre}, the opposite regime $p/M\gg1$ can be under control of the probe approximation. The reason is that the length scale associated to anti-D3 backreaction grows as the fourth root of $g_s p$ whereas the length scale of the A-cycle grows as the square root of $g_sM$. Hence there is a consistent limit in which the anti-brane backreaction is small from the point of view of the $S^3$ despite having $p\gg M$. 

The flux cascades that arise in these set-ups can be effectively interpreted as a model of axion monodromy. As the NS5-brane bounces back and forth on the $S^3$ the flux quanta $K$ drop, the anti-branes are annihilated and the SUSY breaking energy lowers as a perturbative potential superimposed with small wiggles. This is reminiscent of standard axion monodromy models~\cite{Silverstein:2008sg, McAllister:2008hb, McAllister:2014mpa, Marchesano:2014mla, Blumenhagen_2014, Hebecker_2014}. In fact the KPV cascade seems to correspond to a stringy embedding of the flux unwinding scenario envisaged in \cite{DAmico:2012wal}. 

Unfortunately, turning this brane-flux decay cascade into a viable inflationary scenario with the right magnitude of the CMB power spectrum requires extremely large tadpole numbers~\cite{Gautason:2016cyp}. These could very well be impossible to achieve. We may nevertheless set aside the construction of viable phenomenological scenarios. Brane-flux annihilation cascades with reasonable tadpoles provide an interesting set-up to study conceptual questions, in particular whether effective potentials can be constructed and controlled over large (trans-Planckian) field ranges.  As we will show next, the conifold modulus plays a crucial role: its associated instability imposes severe constraints on the maximal amount of anti-branes that a stable warped conifold admits, and hence on how far the $\psi$-potential can be extended. 

Before entering the details of the computation we list several constraints that the scenario is subject to. This should help the reader keep in mind how non-trivial it is to achieve multiple monodromies in the first place:
\begin{itemize}
\item Couplings, curvature scales and inverse length scales of cycles should be small.
\item The length scale associated to anti-D3 (NS5) backreaction, $(g_sp)^{1/4}l_s$ should be small compared to the linear size of the A-cycle $(g_sM)^{1/2}l_s$.
\item As the $\psi$-modulus traverses the field range considered, other moduli should not get destabilised. In particular, the volume modulus should not shift too much. In other words; the moduli stabilisation scenario, which generically features light K\"ahler moduli, should not get upset. The particular form of this constraint will depend on the mechanism of moduli stabilisation, but will typically require the SUSY breaking ingredients to be sufficiently warped down. This will be reflected in turn in an upper bound on the values that can be probed by the conifold modulus $|Z|$. We should note here that this constraint may be more difficult to satisfy in some set-ups (e.g.~the KKLT scenario) than in others (e.g.~the Large Volume Scenario). Our analysis will be general and independent of the particular mechanism of K\"ahler moduli stabilisation.
\item The throat volume should be smaller than the overall internal volume as a minimal consistency condition. Again, this is a constraint that depends strongly on the method of K\"ahler moduli stabilisation, towards which our analysis is agnostic.
\end{itemize}
We will focus in the following on the requirement that the conifold modulus should remain stable throughout the brane-flux cascade. This constraint has not been studied before. We will find that it has a crucial effect in restricting the maximal amount of monodromies that can be reached. 

\subsection{Anti-brane annihilation}

There is another subtlety that we should clarify before analysing the monodromy potential. Up to now, we have kept the three-form flux fixed, given by the (large) parameter $K$. This was a sufficiently good approximation for our analysis of metastable vacua, in which no brane-flux annihilation occurred, i.e. $\psi<\pi$. In the following, we will be interested in processes in which brane-flux annihilation takes place repeatedly. After each monodromy $\psi\to\psi+\pi$, $M$ units of anti-D3-branes are annihilated $Q_\text{D3}\to Q_\text{D3}+M$. This is reflected in the fact that all the dependence on $p$ of previous formulae is in terms of the combination $p-\frac{M\psi}{\pi}\,+\text{(periodic modulations)}$. Hence, in our conventions, $p$ corresponds to the initial number of anti-D3s before annihilation takes place. In a similar fashion, the total amount of three-form flux undergoes a shift $K_\text{tot}\to K_\text{tot}-1$ after each monodromy, as dictated by the tadpole condition $K_\text{tot}M+Q_\text{D3}=0$. This means that $K_\text{tot}$ depends on $\psi$ as $K_\text{tot}=K+p/M-\psi/\pi\,+\text{(periodic modulations)}$. With this convention, $K$ corresponds to the flux quantum present at the end of the cascade $\psi=\psi_f=\pi p/M$, when the $p$ units of $\overline{\text{D3}}$-branes have disappeared. All previous expressions in this paper hold upon the replacement $K\to K_\text{tot}(\psi)$.

Having clarified this, we can easily estimate the maximum number of monodromies and the associated maximum field range $\Delta \psi$ covered by the process of anti-brane annihilation. In the limit $\frac{p}{M} \gg 1$, the oscillatory behaviour in the $v_\text{NS5}$ factor of the potential is subdominant. $v_\text{NS5}$ hence becomes effectively linear in $\psi$, like in other axion monodromy models~\cite{Gautason:2016cyp} :
\begin{align}
\label{eq:vns5_linear}
	v_\text{NS5} \approx |\tilde{\psi}| + \tilde{\psi} = \left\{\begin{array}{ll} 2 \tilde{\psi} &\text{ for }\tilde{\psi} > 0 \\
		0 &\text{ for } \tilde{\psi} < 0 \end{array} \right.
\end{align}
where we have defined $\tilde{\psi} \equiv \pi p /M - \psi$ as $\psi$ always appears in this combination ($M\tilde{\psi}/\pi$ can be thought of as the total amount of anti-brane charge present for a given value of $\psi$). The field $\psi$ rolls down a potential with small modulations across a certain field range, serving potentially as an inflaton candidate. We can see that the bound~\eqref{eq:vns5_bound} resulting from the conifold instability immediately constrains the maximum range in the SUSY-breaking regime (c.f. Figure~\ref{fig:boundPot}):
\begin{equation}\label{DeltaPsi}
	\Delta \psi =\Delta \tilde{\psi} < \Delta \tilde{\psi}_\text{max}=\frac{9 g_s M}{64 c' c''}\approx \,0.07 g_s M\,.
\end{equation}

\begin{figure}[h]
\centering
	\begin{tikzpicture}
    	\node[anchor=south west,inner sep=0] (image) at (0,0) {\includegraphics[width=0.6\textwidth]{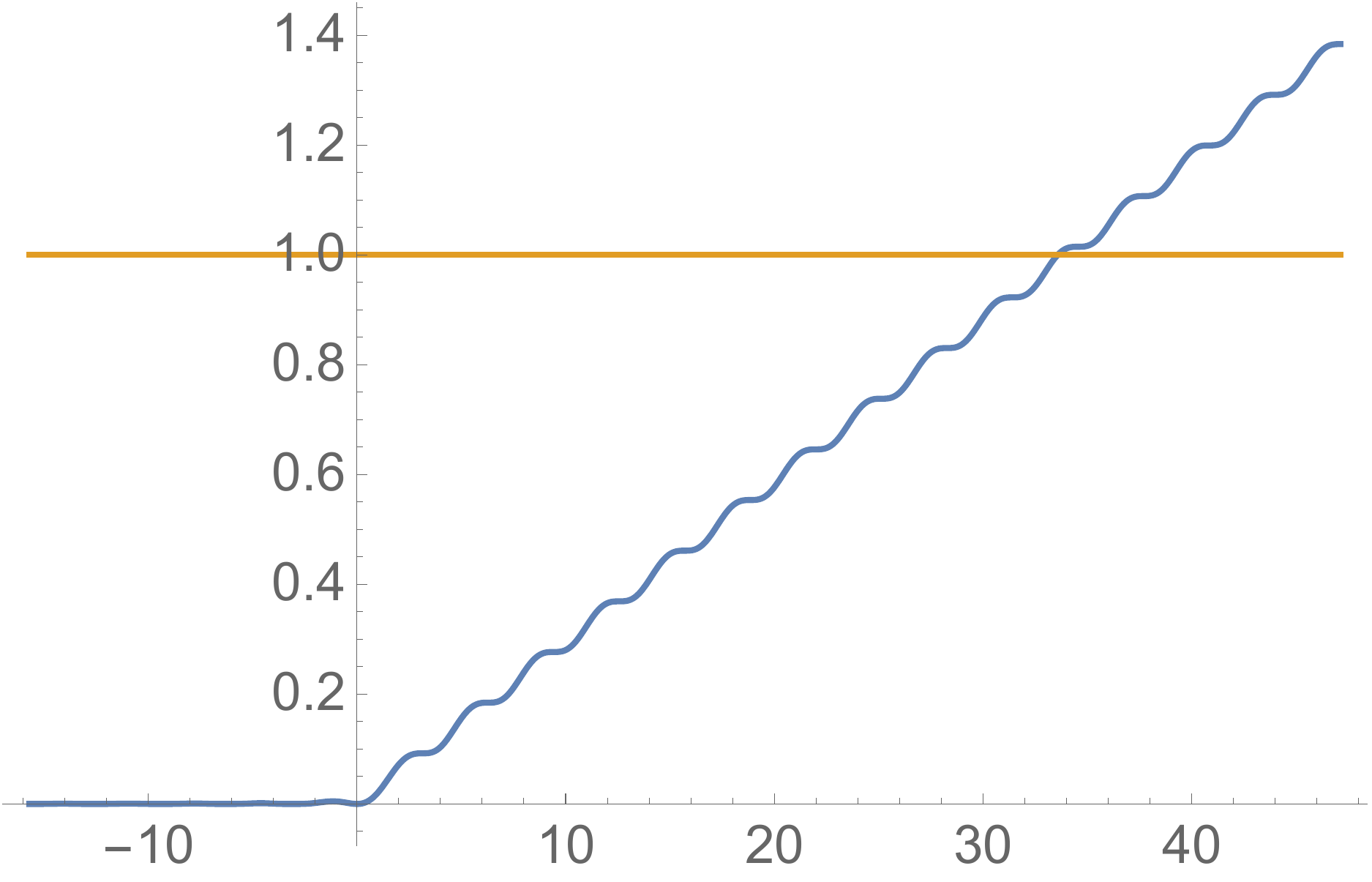}};
    \begin{scope}[
        x={(image.south east)},
        y={(image.north west)}
    ]
        \node [black] at (-0.02,1) {\large $\frac{32 c'c''}{9 g_s M}v_\text{NS5}$};
        \node [black] at (1.05,0.05) {\large $\tilde{\psi}$};
    \end{scope}
    \end{tikzpicture}
	\caption{The bound \eqref{eq:vns5_bound} constrains the allowed field range of $\tilde{\psi}$. The maximum field range is obtained where the yellow and blue line meet, here plotted for $g_s M = 500$.}
	\label{fig:boundPot}
\end{figure}

Two comments are in order:
\begin{itemize}
\item While the maximum allowed field range $\Delta \tilde{\psi}$ is determined by~$v_\text{NS5}$ as in~\eqref{DeltaPsi}, the dependence on $\tilde{\psi}$ of the total potential is more complicated. It should read:
\be\label{eq:totpot}
V_\text{tot}=V_\text{KS}+V_\text{NS5}=\frac{M_{p}^4|Z|^{4/3}}{\sigma^2}\left[\frac{1}{2c'}\left|\frac{1}{2\pi}\log Z+\frac{K_\text{tot}(\tilde{\psi})}{g_s M}\right|^2+\frac{c''}{4\pi^2}\frac{1}{g_s M}v_{NS5}(\tilde{\psi})\right]\,.
\ee
There is an explicit $\tilde{\psi}$-dependence in the $V_\text{KS}$ contribution through $K_\text{tot}(\tilde{\psi})$. Furthermore, there is an implicit dependence via the $Z$-modulus. Indeed, as seen from~\eqref{eq:minimumZ}, the minimum of $Z$ depends exponentially on the total 3-form flux $K_\text{tot}$, which shifts by an amount $\Delta \tilde{\psi}$ along the trajectory. Freezing the value of $Z$ at some fixed value, say at its minimum~\eqref{eq:minimumZ} with $\tilde{\psi}=0$, would put $Z$ far off-shell after the anti-branes are generated. It would correspond to freezing the warp factor throughout the cascade. Not only is this not the trajectory followed dynamically, but the used $Z$-potential  may not be fully trusted far away from its minimum in any case. 

Hence, we are dealing effectively with a complicated two modulus $(\tilde{\psi},Z)$ problem. We can simplify it significantly if we assume that the $Z$-modulus will change with $\tilde{\psi}$ by tracing its consecutive minima, as dictated by eq.~\eqref{eq:minimumZ}, taken as a function of $\tilde{\psi}$ (recall that we should replace $K\to K_\text{tot}(\tilde{\psi})= K + \tilde{\psi}/\pi = K+p/M-\psi/\pi$ in that equation). The corresponding total potential $V_\text{tot}(\tilde{\psi})$ is plotted in Figure~\ref{fig:fullKshiftedpot}. 
\begin{figure}[h!]
\centering
	\begin{tikzpicture}
    	\node[anchor=south west,inner sep=0] (image) at (0,0) {\includegraphics[width=0.6\textwidth]{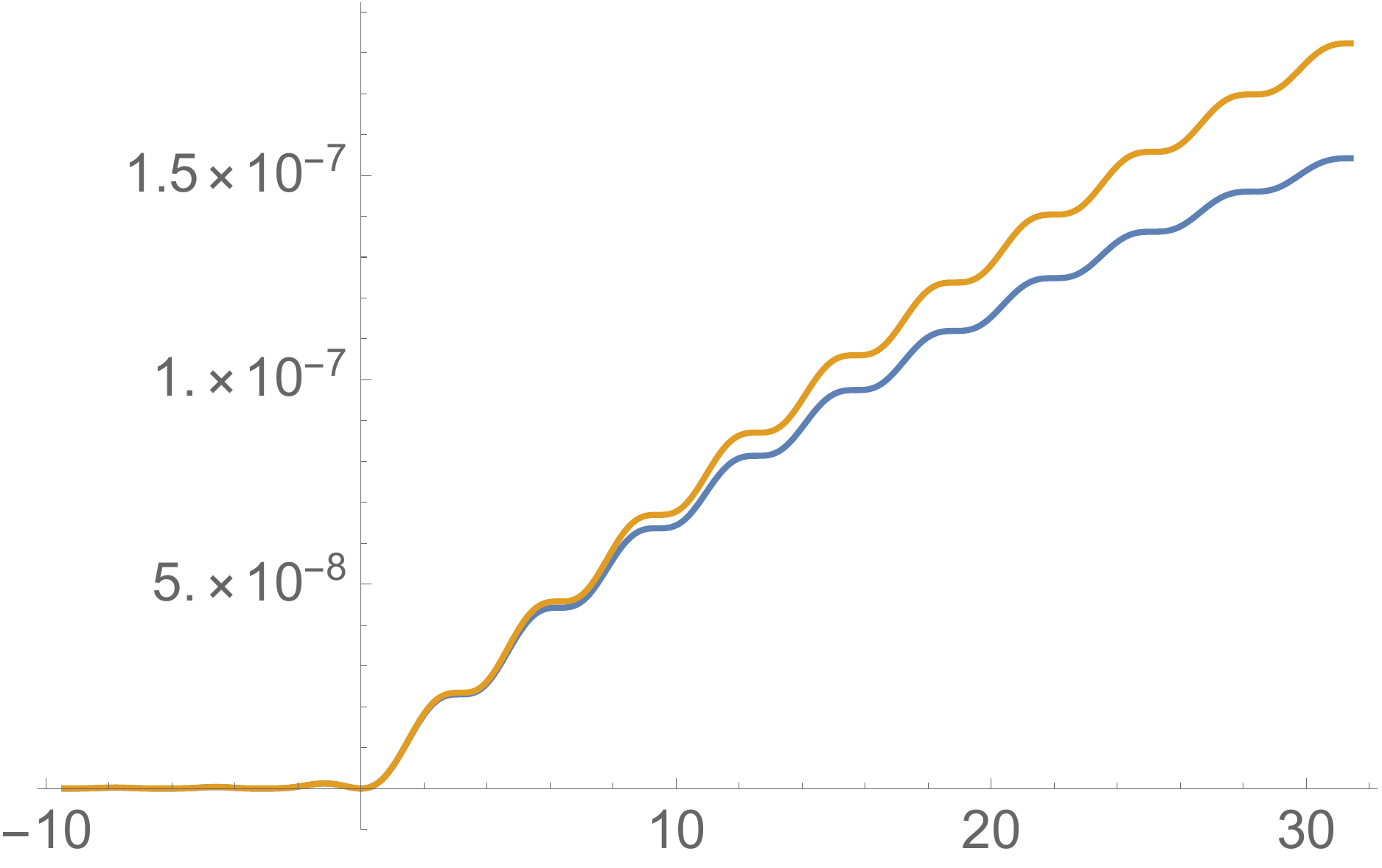}};
    \begin{scope}[
        x={(image.south east)},
        y={(image.north west)}
    ]
        \node [black] at (-0.01,1) {\large $M_\text{Pl}^{-4} \sigma^2 V_\text{tot}(\tilde{\psi})$};
        \node [black] at (1.05,0.05) {\large $\tilde{\psi}$};
    \end{scope}
\end{tikzpicture}
	\caption{The shift $K \to K_\text{tot}$ changes the potential from the yellow to the blue curve. However, the maximum field distance $\Delta \tilde{\psi}_\text{max}$ is not affected by the change since it does not depend on $K$. The potentials are plotted here for $g_s M = 500$ and $K=600$.}
	\label{fig:fullKshiftedpot}
\end{figure}

\item The field $\psi$ whose maximum range is bounded as in~\eqref{DeltaPsi} is not canonically normalised. The invariant field displacement $\Delta\phi$ should be obtained by integrating\\ $\dd \phi \approx \sqrt{g_{ZZ} \dd Z^2 + g_{\psi\psi}\dd\psi^2}$ along the trajectory (here we neglect possible kinetic mixings between $Z$ and $\psi$). This is a rather complicated problem. In the following, we make several conservative approximations to establish an upper bound on the maximum field range.
\end{itemize}

As explained above, we are interested in a field trajectory where $Z$ remains in its minimum, constantly adjusting its value when $\psi$ moves. The canonical field distance will be bounded by 
\begin{align}
\label{eq:distance_estimate}
	& \Delta \phi \leq \Delta_Z\phi + \Delta_\psi \phi, \\
	& \text{with}\quad \Delta_Z \phi \equiv M_\text{Pl}\left|\int_{|Z|_{\tilde{\psi}_i}}^{|Z|_{\tilde{\psi}_f}}\dd Z \sqrt{g_{ZZ}}\right|, \quad \Delta_\psi \phi \equiv M_\text{Pl}\left|\int_{\tilde{\psi}_i}^{\tilde{\psi}_f}\dd \tilde{\psi}\sqrt{g_{\psi\psi}} \right|.
\end{align}
The second contribution is restricted by~\eqref{DeltaPsi}. Then, we take $\tilde{\psi}_i = 0$ and $\tilde{\psi}_f = \Delta \tilde{\psi}_\text{max}$.

As a first step, we compute the second term in \eqref{eq:distance_estimate}, i.e. $\Delta_\psi \phi$.
The kinetic term for $\psi$ was computed in \cite{Kachru:2002gs}, which we briefly reproduce here. In what follows we rely strongly on the formulas in \cite{Kachru:2002gs, DeWolfe:2004qx,  Blumenhagen:2019qcg}.\footnote{We have corrected a power of $z$ in text below equation (5) in \cite{Kachru:2002gs} which seems to be a typo.} 

The induced metric on the NS5 at the tip of the throat (in 10d string frame) reads
\be
\dd s^2 = e^{2A(0)} \left[- \left(1-b_0^2 \,g_s\, M \alpha' e^{-2A(0)}\dot{\psi}^2\right) \dd t^2 + a^2(t) \dd x^i \dd x^i \right] + b_0^2\, g_s\, M \alpha' (\sin^2 \psi) \dd\Omega_2^2\,,
\ee
allowing for a cosmological scale factor $a(t)$. The 4d coordinates are kept dimensionful whereas the $S^3$ coordinates (like $\psi$) are dimensionless. The kinetic term computed in \cite{Kachru:2002gs} then becomes:\footnote{We expand here $\sqrt{1 - \alpha\dot{\psi}^2}$ as $1- \frac{1}{2}\alpha\dot{\psi}^2$ and neglect periodic modulation effects in the anti-brane charge $\frac{\pi p}{M}-\psi+\ldots$. As mentioned, the latter is a good approximation in the regime of interest $p\gg M$. } 
\begin{align}
S &= \int \dd t \dd^3x \,a(t)^3\,\left[\frac{\mu_3 M}{\pi g_s}b_0^2 \,g_s\, M \alpha' e^{2A(0)} \left(\frac{\pi p}{M}-\psi\right)\dot{\psi}^2\right]\nonumber\\
&\approx M_{\text{Pl}}^2\, \int \dd t \dd^3x \,a(t)^3\,\left[0.01\, \frac{M|Z|^{2/3}}{\sigma}  \left(\frac{\pi p}{M}-\psi\right)\dot{\psi}^2\right].
\end{align}
The integrand in the second term of \eqref{eq:distance_estimate} field can then be expressed as
\be
\sqrt{g_{\psi\psi}} = 0.1 \, \frac{M^{1/2}|Z|^{1/3}}{\sigma^{1/2}}\,\sqrt{\tilde{\psi}}\,.
\ee
As explained, in these expressions we should think of the warp factor in terms of the $Z$-modulus, which in turn should be considered a function of $\psi$. 

The contribution $\Delta_\psi \phi$ for our bound is given by  
\be
\label{eq:distancepsiapprox}
\Delta_\psi \phi  = M_\text{Pl}\left|\int^{\Delta\tilde{\psi}_\text{max}}_0 \dd\tilde{\psi}\: \sqrt{g_{\psi \psi}}\right|,
\ee
where the bound on the field range $\Delta \tilde{\psi}_\text{max}$ was determined in~\eqref{DeltaPsi}. Rather than evaluating the complicated integral explicitly, we will make a conservative approximation by freezing $|Z|_{\tilde{\psi}}$ at its maximal value in the integration domain. This happens when $\tilde{\psi}= 0$, where $|Z|_{\tilde{\psi}=0}= \exp(-2\pi K /g_s \, M)$. This approximation allows an analytical solution to~\eqref{eq:distancepsiapprox}:
\begin{align}
	\Delta_\psi \phi <& 0.1 \,M_{\text{Pl}}\, \frac{M^{1/2}}{\sigma^{1/2}} |Z|^{\frac{1}{3}}_{\tilde{\psi}=0}\, \int_{0}^{\Delta\tilde{\psi}_\text{max}} \dd\tilde{\psi}\:\sqrt{\tilde{\psi}} \approx 0.1 M_{\text{Pl}}\sqrt{\frac{M}{\sigma}}e^{-\frac{2\pi K}{3g_s M}}(\Delta \tilde{\psi}_\text{max})^{3/2}\nonumber\\
	&\approx 0.001\,M_{\text{Pl}} \frac{(g_s M)^2}{{\cal V}_s^{1/3}}\,e^{-\frac{2\pi K}{3g_s M}}\, ,\label{eq:distancebound}
\end{align}
where we introduced $\mathcal{V}_s^{2/3} \equiv g_s \sigma = \mathcal{V}_w^{2/3} g_s$, the string frame volume in string units assumed to be large in order to trust effective field theory.

We now calculate an upper bound for $\Delta_Z \phi$. We notice from \eqref{eq:kahlerpot} with $g_{ZZ} = \partial_Z \partial_{\bar{Z}} \mathcal{K}$ that
\begin{align}
\nonumber
	\Delta_Z \phi &= M_\text{Pl} \left|\int_{|Z|_{\tilde{\psi}_i}}^{|Z|_{\tilde{\psi}_f}}\dd Z \sqrt{g_{ZZ}} \right|\\
\nonumber
	&=M_\text{Pl} \left|\int_{|Z|_{\tilde{\psi}_i}}^{|Z|_{\tilde{\psi}_f}} \dd Z \left(\frac{1}{\mathcal{V}_s^{1/3}} \sqrt{\frac{c'}{8}} g_s M |Z|^{-2/3}\right)\right|\\
	&= M_\text{Pl}\frac{3}{\mathcal{V}_s^{1/3}} \sqrt{\frac{c'}{8}} g_s M \left
	(|Z|^{1/3}_{\tilde{\psi}_i} - |Z|^{1/3}_{\tilde{\psi}_f}\right)\,.
\end{align}
Again, we notice that $Z$ takes its maximum value at $\tilde{\psi}_i = 0$ and we conclude that 
\be\label{eq:distancebound2}
	\Delta_Z \phi < M_\text{Pl} \frac{g_s M}{\mathcal{V}_s^{1/3}} e^{-\frac{2\pi K}{3g_s M}}\,.
\ee
We see that this contribution has a factor $g_s M$ less than $\Delta_\psi \phi$.

Overall, we find a bound on the total canonical field distance of the form:
\be\label{eq:totaldistancebound}
\Delta\phi<M_\text{Pl} \,\frac{g_s M}{\mathcal{V}_s^{1/3}} \,e^{-\frac{2\pi K}{3g_s M}}\left(1+10^{-3}g_s M\right)
\ee
The maximum field displacement is suppressed both by the volume and the exponential hierarchy $Z^{1/3}$. Interestingly these are two parameters that control the approximations made in the compactification procedure.\footnote{According to \cite{DeWolfe:2004qx} the string frame volume of the throat, $\mathcal{V}_{th}$ was argued to be
$\mathcal{V}_{th}^{1/3} = \alpha (g_sMK)^{1/2} > \alpha Mg_s$, where $\alpha$ is some numerical constant that can not really be computed but was guessed to be upperbounded by the number $\sqrt{3}\pi^{3/2}$ in \cite{DeWolfe:2004qx}.} One might be tempted to extend the possible field range by compensating the exponential warping and the volume suppression by a large polynomial dependence in $(g_s\,M)$, i.e. by imposing  $g_s\,M + 0.001\,(g_s\,M)^2\gg {\cal V}_s^{1/3} e^{2\pi K/3g_sM}$. We deem this possibility as highly unlikely: it requires fairly large tadpole numbers and a regime of poorly controlled approximations, where neither ${\cal V}_s$ nor $K/g_sM$ can be made significantly large (i.e. a regime of small volume and mild warping).

\subsection{Brane creation from fluxes}}

We have argued that the (canonically normalised) field distance traversed in the process of $\overline{D3}$-brane annihilation is exponentially suppressed and cannot be made transplanckian in a controlled regime. The reason being that the conifold modulus becomes destabilised upon inclusion of a large enough number of $\overline{\text{D3}}$-branes. This result is in line with the general spirit of the Swampland program that postulates pathologies upon such large field displacements (c.f.~\cite{Baume:2016psm,Valenzuela:2016yny,Blumenhagen:2017cxt,Blumenhagen:2018hsh,Buratti:2018xjt,Scalisi:2018eaz}). One must however be careful before drawing too strong conclusions in this respect. 

Swampland constraints apply to the total field distance within a given effective field theory. In our set-up, the three-form flux $K$ can still be transferred into D3-charge even after all anti-branes have disappeared. This process can still be described as mediated by NS5-branes, and hence by the field $\tilde{\psi}$ in the regime $\tilde{\psi} < 0$. The maximum field displacement is determined by the maximum flux $K$ available, which is itself constrained by the tadpole condition
\begin{align}
	\frac{\chi(\Sigma)}{24} =& K_\text{tot}M -p_\text{tot} - \frac{1}{4} N _\text{O3}
	=\left(K+\frac{\tilde{\psi}}{\pi}\right)M -\frac{M \tilde{\psi}}{\pi} - \frac{1}{4} N _\text{O3}
\end{align}
where the explicit $\tilde{\psi}$ dependence reflects the change in flux and D3-charge mediated by $\tilde{\psi}$. 

One may wonder whether large total field ranges can be found if there exist compactification spaces that support large enough fluxes. This is quite unlikely in our opinion. One should notice that as long as there is significant warping  the canonically normalised field displacement will be exponentially suppressed, as in~\eqref{eq:distancebound}. Warping is crucial in keeping control of the effective field theory we have been working with. Indeed, warping is the only reason why brane-flux annihilation mediated by NS5-branes winding around internal cycles are not immediately highly energetic processes out of the regime of the EFT. Hence, a large (non-exponentially-suppressed) field displacement requires leaving the regime of control of the theory where $\tilde{\psi}$ is a light field. This is reflected in the potential $V_\text{tot}(\tilde{\psi})$, eq.~\eqref{eq:totpot}, in particular through the implicit $Z(\tilde{\psi})$ dependence. 

It is also worth mentioning that field displacements along $\psi$ seem to be related to the appearance of light modes, a feature generically postulated by the Swampland Distance Conjecture. In the regime of large positive $\tilde{\psi}$, i.e. when a large number of anti-branes are present and the throat is highly warped, there appear light KK geometric modes. Unlike in typical situations addressed by the distance conjecture, the mass of these states does not seem to scale exponentially with the canonically normalised field distance~\cite{Blumenhagen:2019qcg}. As one moves towards large negative $\tilde{\psi}$, i.e. as the throat become shorter, these states become heavier and ultimately disappear from the low energy EFT. Other fields would at the same time become light, e.g. D-branes wrapping the dual B-cycle, which gets shorter  as $\tilde{\psi}$ shrinks (and thus $\psi$ grows) (c.f. section~\ref{Sec:5branes}). 

Overall, there seem to be suggestive connections between these considerations and the Swampland program (constraints on field ranges, appearance and disappearance of light fields, etc.) but the interpretation is obscured due to the presence of warping and axion monodromies. These issues are clearly worth further studies.

\section{Discussion}\label{Sec:discussion}

In this note we have argued that axion monodromies related to brane-flux transitions down warped throats come with exponentially suppressed field ranges as given by equation~\eqref{eq:totaldistancebound}. Such axion monodromies could potentially be interesting for embedding the ``unwinding flux" inflation scenario \cite{DAmico:2012wal} into string theory as claimed in \cite{Gautason:2016cyp, DiazDorronsoro:2017qre}. However, our findings point at a concrete challenge for such type of models, when a super-Planckian field range is required to deliver a minimum amount of inflation (note this is the case for a linear potential). The reason for the bounded field range~\eqref{eq:totaldistancebound} is the dynamics of the conifold modulus, which was not taken into account in the analysis of \cite{Gautason:2016cyp, DiazDorronsoro:2017qre}. The dynamics of the conifold modulus was recently used to bound meta-stable dS vacua from anti-brane uplifts \cite{Bena:2018fqc}, although, as discussed, the concrete bounds on fluxes were already enforced from other considerations such as brane-flux stability and having a throat tip that is within the supergravity approximation. Nevertheless, the suggestion of~\cite{Bena:2018fqc} does impact strongly certain models of axion monodromy, as we showed. Another interesting use of the conifold modulus is in providing a consistent higher dimensional picture of supersymmetry breaking~\cite{Randall:2019ent}.

Setting aside concrete applications, it is of theoretical importance to understand whether axion monodromies can accommodate large field ranges in controlled regimes and hence challenge Swampland conjectures. Our findings suggest that certain axion monodromy set-ups obtained from unwinding fluxes cannot. 
It would be interesting to investigate the relation of our results to other models of  axion-monodromy inflation that involve flux transitions and/or warped-down axions (e.g.~\cite{McAllister:2008hb, Franco:2014hsa, Ibanez:2014swa, Retolaza:2015sta, Buratti:2018xjt, Hebecker:2018yxs, Kim:2018vgz}), and whether they are also affected by the stability of the conifold modulus.

\section*{Acknowledgements}
We would like to thank Iosif Bena, Mariana Gra\~na, Arthur Hebecker, Luca Martucci, Jakob Moritz, Lisa Randall and Filip Sevenants for useful discussions, and the KITP for hospitality during the ``String Swampland and Quantum Gravity Constraints on Effective Theories'' program. VVH is supported by grant nr.~1185120N of the Research Foundation - Flanders (FWO).  MS acknowledges partial support by the Research Foundation - Flanders (FWO) and the European Union's Horizon 2020 research and innovation programme under the Marie Sk{\l}odowska-Curie grant agreement No. 665501. This research was supported in part by the National Science Foundation under Grant No. NSF PHY-1748958.

\appendix

\section{Mass ratio of the conifold and brane-flux moduli}
\begin{figure}[h!]
\centering
	\begin{tikzpicture}
    	\node[anchor=south west,inner sep=0] (image) at (0,0) {\includegraphics[width=0.6\textwidth]{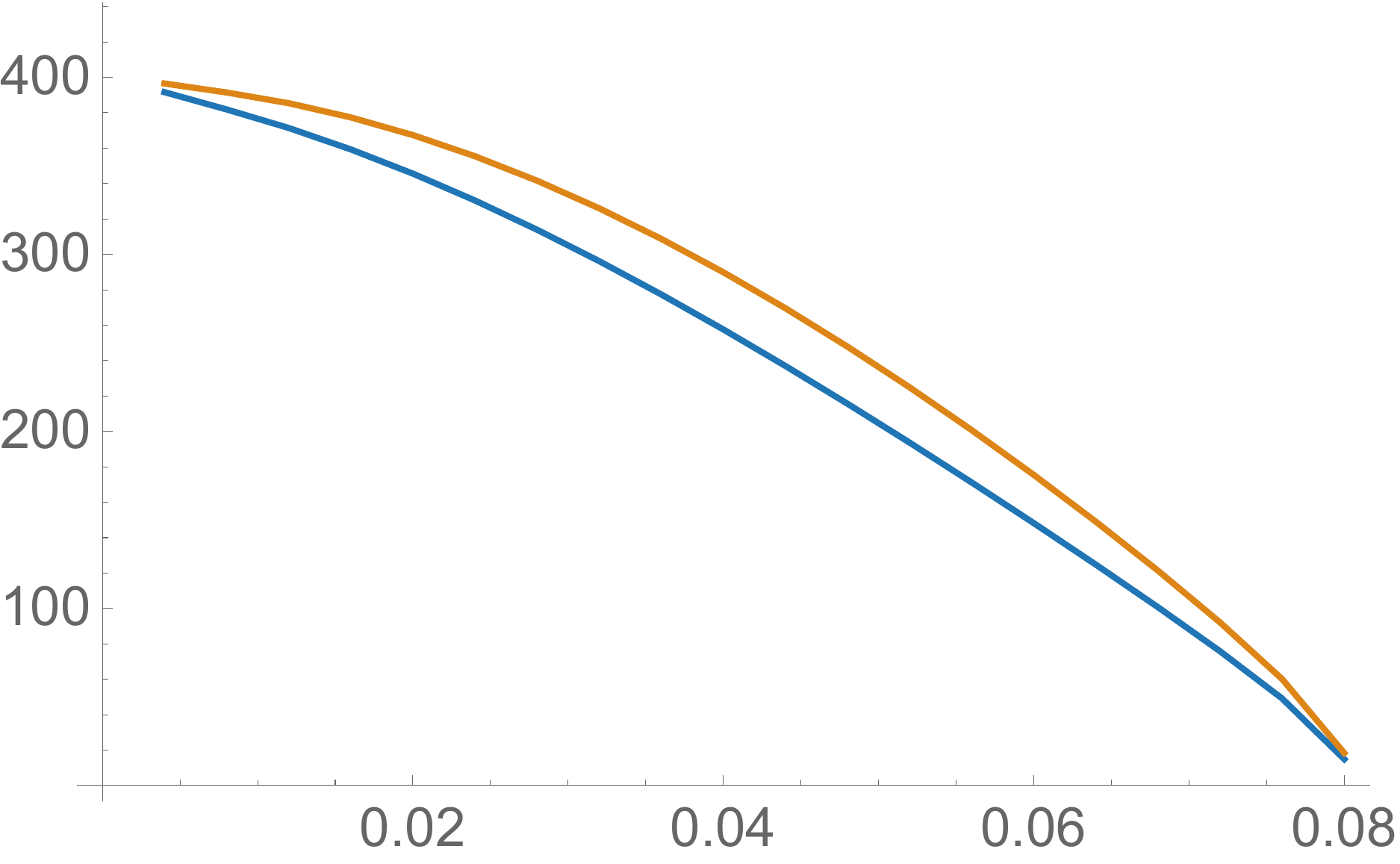}};
    \begin{scope}[
        x={(image.south east)},
        y={(image.north west)}
    ]
        \node [black] at (0.02,1.05) {\large ${m_{\psi}^2}/{m_Z^2}$};
        \node [black] at (1.07,0.05) {\large $p/M$};
    \end{scope}
\end{tikzpicture}
	\caption{The mass ratio ${m_{\psi}^2}/{m_Z^2}$, plotted here for $g_s M =10$ (blue) and $g_s M =100$ (yellow), decreases for growing $p/M$, until the KPV vacuum is lost at $p/M\approx 0.08$. The mass ratio is thus of $O(10)-O(100)$, $\psi$ being the heavier modulus. This does not change qualitatively for different values of $g_sM$.}
	\label{fig:massratio}
\end{figure}

In order to compute the masses of the scalars, we need their potential and their kinetic terms. We work with two fields, $Z$ and $\psi$.
From the KPV analysis we find that the action for the NS5-brane is
\begin{equation}
S_\text{NS5} = \int \dd^4 x \sqrt{-\tilde{g}} \frac{M_\text{Pl}^4}{2\pi} \frac{M e^{4A}}{\pi \sigma^3} \left((v_\text{NS5}-U)\sqrt{1-b_0^2 g_s M \alpha'e^{-2A}\dot{\psi}^2} + U\right),
\end{equation} 
where $v_\text{NS5}$ and $U$ were defined in~\eqref{eq:vns5andU}.
We now expand the square root for small velocities $\dot{\psi}$. In doing so, we find
\begin{equation}
S_\text{NS5} = \int \dd^4 x \sqrt{-\tilde{g}} \left(\frac{1}{2} M_\text{Pl}^2g_{\psi\psi} \dot{\psi}^2 - V_\text{NS5}\right), \quad g_{\psi\psi} = \frac{1}{(2\pi)^2}\frac{2 g_s M |Z|^\frac{2}{3}}{\pi \mathcal{V}_s^{1/3}} \left(\frac{2}{3}\right)^\frac{1}{3}(v_\text{NS5}-U).
\end{equation}
The kinetic term for the conifold modulus can be obtained from the K\"ahler potential \eqref{eq:kahlerpot}. To leading order in $Z$ and large volume, we find\footnote{We ignore the off-diagonal component $g_{Z\rho}$ as it is subleading in small-$Z$ and large volume.}
\begin{equation}
g_{ZZ} = \partial_{Z}\partial_{\bar{Z}}\mathcal{K} = \frac{c'}{8} \frac{(g_s M)^2}{\mathcal{V}_s^{2/3}} |Z|^{-\frac{4}{3}}.
\end{equation}
The Hessian of the total potential $V_\text{tot}$ is diagonal for $Z$ and $\psi$ due to the structure of the potential. The kinetic terms for these fields are also diagonal. Therefore, their mass ratio is given by
\be
	\frac{m_{\psi}^2}{m_Z^2} = \frac{g_{\psi\psi}^{-1}\: \partial_\psi^2 V_\text{tot}}{g_{ZZ}^{-1} \: \partial_Z^2 V_\text{tot}}\Biggr|_{|Z|_0, \, \psi_0},
\ee
where it has to be evaluated at the minimum ($|Z|_0, \, \psi_0$) of the potential. We plotted this mass ratio as a function of $p/M$ in Figure~\ref{fig:massratio}. The ratio appears to be independent of the flux number $K$. It is of order $O(10)-O(100)$ an is qualitatively independent of $g_s M$.

\bibliographystyle{utphys}
\bibliography{refs}

\end{document}